\documentclass{jaa}

\usepackage{graphicx}

\begin{document}\sloppy

\title{Can a regular black hole be observationally distinguished from singular black holes as spinning lens partner in PSR-BH binaries?}


\author{G.Y. TULEGANOVA\textsuperscript{1}, R.KH. KARIMOV\textsuperscript{1}, R.N. IZMAILOV\textsuperscript{1,2,*} and K.K. NANDI\textsuperscript{1,3,4}}
\affilOne{\textsuperscript{1}Zel'dovich International Center for Astrophysics, Bashkir State Pedagogical University, 3A, October Revolution Street, Ufa 450008, RB, Russia.\\}
\affilTwo{\textsuperscript{2}Institute of Molecule and Crystal Physics, Ufa Federal Research Centre, Russian Academy of Sciences, Prospekt Oktyabrya 151, Ufa 450075, Russia.\\}
\affilThree{\textsuperscript{3}Department of Physics \& Astronomy, Bashkir State University, 47A, Lenin Street, Sterlitamak 453103, RB, Russia.\\}
\affilFour{\textsuperscript{4}High Energy Cosmic Ray Research Center, University of North Bengal, Darjeeling 734 013, WB, India.}


\twocolumn[{

\maketitle

\corres{izmailov.ramil@gmail.com}

\msinfo{20 July 2020}{}

\begin{abstract}
To answer the question posed in the title, we consider a novel diagnostic, viz., the difference in the times of arrival (TOA) at the observer of two light rays that simultaneously emanate from a source behind a spinning lens and pass by either side of the lens to reach the observer. This is completely different from the usual Shapiro gravitational time delay, where only one onward light ray is reflected back to the observer. The TOA essentially samples the frame dragging caused by the spinning lens, apart from other lens parameters. Assuming a charged \textit{regular} Ay\'{o}n-Beato and Garc\'{\i}a black hole as the spinning lens partner in some typical astrophysical PSR-BH binaries, which provide the best laboratory for testing the TOA effect, we theoretically study how the prediction depends on the gyromagnetic ratio $\left(Q/M\right)$ and how it compares with those when the role of spinning lens partner is played by the centrally \textit{singular} Kerr-Newman and Kerr black holes. The numerical estimates for two illustrative binary lens systems show $\mu$sec level delay at the zeroth order, which should be measurable. However, the TOA predictions under thin-lens approximation are shown to differ only at third or higher orders of smallness indicating that the regular and singular black holes \textit{cannot} be observationally distinguished despite significant qualitative differences existing among them.
\end{abstract}


}]


\doinum{12.3456/s78910-011-012-3}
\artcitid{\#\#\#\#}
\volnum{000}
\year{0000}
\pgrange{1--}
\setcounter{page}{1}
\lp{1}

\section{Introduction}

The Ay\'{o}n-Beato and Garc\'{\i}a black hole (AGBH) is a new class of the static spherically symmetric regular exact solutions of General Relativity (GR) coupled to nonlinear electrodynamics (Ay\'{o}n-Beato \& Garc\'{\i}a 1998). See also (Bronnikov 2000) for a very illuminating discussion of the solution. The aim in (Ay\'{o}n-Beato \& Garc\'{\i}a 1998) was to solve what the authors called one of the "mysterious" properties of a black hole - the central singularity -- which is usually considered as one of defects of GR. The AGBH solution of GR indeed removes the defect -- the singularity disappears on account of the coupling of gravity to nonlinear electrodynamics. One may imagine that at each point in spacetime the interaction of gravitation and nonlinear electromagnetism regularizes both the fields. The solution corresponds to a regular BH when the gyromagnetic ratio $\left\vert Q\right\vert /M\leq 0.634$, where $Q$ is the charge and $M$ is the asymptotic ADM mass. The curvature invariants and the electric field are regular everywhere including at the origin.

Manko \& Ruiz (2016) demonstrated that the total energy of electromagnetic field in the static AGBH is equal to only the ADM mass parameter $M$ independent of the charge parameter $Q$. This result supports the original idea of  Born \& Infeld (1934a; 1934b) to use nonlinear electrodynamics for proving the electromagnetic nature of mass. Toshmatov et al. (2015) studied scalar, electromagnetic and gravitational test fields in the AGBH spacetime and showed that damping of the quasinormal modes in regular black hole spacetimes is suppressed in comparison to the case of Schwarzschild black holes. Further, increasing the charge parameter of the regular black holes increases reflection and decreases transmission factor of incident waves for each of the test fields. Nandi et al. (2020) have demonstrated that the famous hoop conjecture (due to Thorne) is not violated in the AGBH spacetime. These are some of the interesting features of the static AGBH.

For our theoretical study, we shall need a spinning AGBH to act as a lens in typical astrophysical binary systems. Using the well known Newman-Janis algorithm (Newman \& Janis 1965), the static, spherically symmetric, charged AGBH has been converted by Toshmatov et al. (2014) into a spinning AGBH that shows a number of interesting properties: It is also regular and the critical value of the electric charge $Q$, for which two horizons merge into one, sufficiently decreases in the presence of non-vanishing spin $a$ of the black hole (Toshmatov et al. 2014). Abdujabbarov et al. (2016) showed that the radius of the shadow cast by the spinning AGBH decreases monotonically and the distortion parameter increases when the value of total mass $M$, rotation parameter $a$ and electric charge $Q$ increase. Along the line of exploring other interesting features, we shall consider the TOA effect under thin-lens approximation. Both the concepts are explained below.

Difference in the times of arrival (TOA) at the observer is a potential new observable, first conceived by Dymnikova (1984, 1986) to our knowledge, that samples the frame dragging effect caused by a spinning lens. Consider a binary system, where a variable light source $S$ (pulsar) orbits a spinning compact lens (BH). As illustrated in Fig.1, suppose two light rays simultaneously emanate from the source $S$ behind a spinning lens $L$ (with mass $M$, spin $a$), pass on either side of the lens to reach the observer at $O$, say, the Earth. The rays will reach at different times at $O$ due to frame dragging caused by the intervening spinning lens. The dragging causes the light path lengths on either side of the lens to differ, shorter on the co-rotating side and longer on the counter-rotating side (Fig.1). The TOA at the observer are \textit{subtracted} to sample the frame dragging effect (absent for a static lens) and so the difference in TOA is also named "relative time delay" by by Laguna \& Wolszczan (1997). We wish to emphasize that the difference in the TOA is \textit{not} the usual Shapiro gravitational time delay (Shapiro 1964) that appears even for a static lens. In this case, a single light ray, sent onward by an observer, passes by an intervening lens of mass $M$ and reflected back to the observer by a distant object in superior conjunction. The total elapsed time at the observer is obtained by \textit{adding} two-way travel times ($M\neq 0$) and compared with similar total elapsed time in the flat space (assuming $M=0$). Shapiro found the former time lapse to be larger than the latter, hence the terminology gravitational time delay\footnote{
There is also a new interesting effect called \textit{gravitational time advancement} first proposed in (Bhadra \& Nandi 2010). Its astrophysical implications are being investigated (Ghosh \& Bhadra 2015; Ghosh, Bhadra \& Mukhopadhyay 2019; Deng \& Xie 2017).}. He found the excess value to be $\delta t_{\text{Shapiro}}^{\text{Sch}}\sim 245\mu $sec with Sun as the gravitating lens, which agrees with observation with remarkable accuracy.

The TOA\ loosely resembles an astrophysical analogue of the quantum Bohm-Aharonov effect. It has first been calculated to the zeroth order by Laguna and Wolszczan (1997) in the Kerr metric for some hypothetical binary systems\footnote{%
We think that the influential work in (Laguna \& Wolszczan 1997) needs some clarification. The starting point, following Dymnikova (1984, 1986), is their \textit{Eq.(1}):
$$t^{\pm}(r,d) = \sqrt{r^{2}-d^{2}}+2M\ln \left( \frac{r+\sqrt{r^{2}-d^{2}}}{d}\right)$$
$$ + M \left(\frac{r-d}{r+d}\right)^{1/2} + \frac{\left(15\pi-8\right) M^{2}}{4d} \mp \frac{4aM}{d},$$
where $d$ is the impact parameter.

The first four are Shapiro terms and the last one is the frame dragging term. The relative time delay of two signals, emitted at point $r_{e}$, traveling in opposite directions around the black hole and arriving at point $r_{0}$ is obtained by subtracting $t^{\pm}$ yielding \textit{Eq.(2)} of (Laguna \& Wolszczan 1997):%
$$\Delta t_{s}=\frac{16aM}{d}.$$

While this is a correct equation to leading order, there is an interpretational issue here. It is implied that the Shapiro terms, which represent \textit{gravitational slowing down of light along radial motion}, have subtracted out. In contradistinction, as stated by Dymnikova (1984), the time difference $\Delta t_{s}$ arises purely due to the \textit{difference in optical paths between the rays}, meaning that the paths are asymmetrical about the optical axis due to different impact parameters $d_{-} = d_{+} + a$. Hence, it is unlikely that the Shapiro slow down along two different paths would completely subtract out, leaving behind just \textit{Eq.(2)}. It is unclear if the magnitude of residual slow down would be negligible compared to $\Delta t_{s}$. A more appropriate leading order delay is our $\Delta t_{1}$ in Eq.(39) that does not involve the slow down residual and is based on realistic finite distance configuration. Further, it yields \textit{Eq.(2)} under the conditions $d_{OL}, d_{LS} \rightarrow \infty$ and thin-lens approximation $(a/d_{+}) << 1$ (see Sec.4 for details). However, the correction to \textit{Eq.(2)} brought about by Eq.(39) is very minute because of the approximations adopted in this paper but the clarification could still be important, when those approximations are waived.}.
A similar, though not exactly the same, type of effect was studied by Datta \& Kapoor (1985), where light rays were assumed to emerge not from a variable source behind the lens but from two diametrically opposite points on a spinning compact astrophysical object itself. A good example of TOA could be the early observation of extremely rapid fluctuations in the brightness of quasar $1525+227$ with characteristic time scale $\sim $ $200$ sec speculated to be caused by a spinning black hole of mass $M\sim 5\times 10^{8}M_{\odot }$ situated between the quasar and the observer (Matilsky, Shr\"{a}der \& Tananbaum 1982). Recently, TOA has been theoretically studied for the Johannsen metric (Izmailov et al. 2019a) as a possible observable diagnostic to test the validity of the so-called "no-hair" conjecture of Penrose. The effect of string parameter on TOA was investigated in (Izmailov et al. 2020). So far no precise experimental data on TOA are available but very accurate future data from suitably identified binaries can constrain the values of the relevant deviation parameters. Such constraints were recently discussed also in the context of other modified gtravity theories (Tuleganova et al. 2020; Tuleganova \& Muhamadieva 2021; Karimov et al. 2018a,b; Kulbakova 2018).

We shall adopt thin-lens approximation (see the details, e.g., in Hartle (2003)), which applies to a situation where the source, lens and observer are all considered as points and the light rays are assumed to travel in straight lines with the deflection taking place only at the lens and it works excellently when the rays travel vast distances compared to the lens size with the impact parameter far larger than the photon sphere. This means that the relevant angles making the quadrilateral in Fig.1 with the optical axis $SLO$ are small. Thus the TOA considered here is essentially a weak field effect due to the thin-lens approximation. A more accurate calculation of TOA should involve integration between two finite points $S$ and $O$ on the optical axis of the exact null geodesic around the spinning lens. Nonetheless, the thin-lens approximation provides a "simple and elegant description of many realistic lensing situations" (Hartle 2003).

The purpose of the present paper is to \textit{quantitatively} evaluate the differences in the TOA in three spacetimes, one is the everywhere-regular spinning charged AGBH and the other two are the well known centrally-singular spinning charged Kerr-Newman black hole (KNBH) and its chargeless limit of Kerr black hole (KBH). We shall then compare the values obtained in the these spacetimes to see if and how much they differ from each other. To do that, we shall derive the influence of the frame dragging as well as other parameters of the spinning lens using the thin-lens approximation. The analytic expression for TOA is applied to some potential PSR-BH binaries assuming that the relevant BH data are independently known.

The paper is organized as follows. In Sec.2, we briefly introduce the spinning AGBH and in Sec.3, we derive the generic formula for the difference in the TOA. In Sec.4, we explicitly calculate the TOA components using the thin-lens approximation. Sec.5 contains numerical values for two binary systems and Sec.6 concludes the paper. We shall take $G=1$, $c=1$ unless specifically restored.

\section{Spinning AGBH}

The dynamics of the theory we are using is governed by the gravitational action $S$ with source of nonlinear electrodynamics (Ay\'{o}n-Beato \& Garc\'{\i}a 1998):
\begin{equation}
S=\int \sqrt{-g}d^{4}x\left[ \frac{1}{16\pi }R-\frac{1}{4\pi }\mathcal{L}(F)\right],
\end{equation}%
where $R$ is the Ricci scalar and $\mathcal{L}$ is a function of $F=\frac{1}{4}F_{\mu\nu}F^{\mu\nu}$, where $F^{\mu\nu}$ is the electromagnetic strength. Using the Legendre transformation
\begin{equation}
\mathcal{H\equiv}2F\mathcal{L}_{F}-\mathcal{L},
\end{equation}%
where $\mathcal{L}_{F}\equiv\frac{\partial\mathcal{L}}{\partial F}$ and $\mathcal{H\equiv H(}P\mathcal{)}$ with $P=\frac{1}{4}P_{\mu\nu}P^{\mu\nu} = F\mathcal{L}_{F}^{2}$. The solution crucially depends on the choice of $\mathcal{H(}P\mathcal{)}$ which for the static AGBH is%
\begin{equation}
\mathcal{H(}P\mathcal{)}=P\frac{\left( 1-3\sqrt{-2Q^{2}P}\right) }{\left( 1+\sqrt{-2Q^{2}P}\right) ^{3}}-\frac{3}{2sQ^{2}}\left( \frac{\sqrt{-2Q^{2}P}}{1+\sqrt{-2Q^{2}P}}\right) ^{5/2},
\end{equation}%
where $s\equiv\frac{\left\vert Q\right\vert}{2M}$ is half of the gyromagnetic parameter, its critical values $s=s_{c}$ define the transition between the regimes of BH horizon and no horizon and the invariant $P$ is negative. The correspondence with linear Maxwell theory is achieved in the weak field ($P\ll 1$) when the usual electromagnetic strength tensor is recovered:
\begin{equation}
F_{\mu\nu}\equiv \left(\frac{\partial\mathcal{H}}{\partial P}\right) P_{\mu\nu}.
\end{equation}

The spherically symmetric, asymptotically flat, static AGBH metric obtained from the Einstein equations is given by (Ay\'{o}n-Beato \& Garc\'{\i}a 1998)
\begin{eqnarray}
d\tau ^{2} &=&-A(r)dt^{2}+\frac{1}{A(r)}dr^{2}+r^{2}(d\theta ^{2}+\sin^{2}\theta d\psi ^{2}), \\
A(r) &=&1-\frac{2Mr^{2}}{(r^{2}+Q^{2})^{3/2}}+\frac{Q^{2}r^{2}}{(r^{2}+Q^{2})^{2}},
\end{eqnarray}%
with the associated asymptotically vanishing nonlinear electric field $E$ given by
\begin{equation}
E=Qr^{4}\left[\frac{r^{2}-5Q^{2}}{(r^{2}+Q^{2})^{4}}+\frac{15}{2}\frac{M}{(r^{2}+Q^{2})^{7/2}}\right].
\end{equation}

Using Janis-Newman algorithm (Newman \& Janis 1965), the spinning AGBH obtained in (Toshmatov et al. 2014) in the Boyer-Lindquist coordinates ($t$, $r$, $\theta$, $\phi$) reads
\begin{equation}
d\tau ^{2}=-g_{tt}dt^{2}-2g_{t\varphi }dtd\phi +g_{rr}dr^{2}+g_{\theta\theta }d\theta ^{2}+g_{\phi \phi }d\phi ^{2},
\end{equation}%
where
\begin{eqnarray}
g_{tt} &=&f(r,\theta), \\
g_{t\varphi } &=&a\sin ^{2}\theta \left[ f(r,\theta )-1\right], \\
g_{rr} &=&\frac{\Sigma }{\Delta _{\text{AGBH}}}, \\
g_{\theta \theta } &=&\Sigma , \\
g_{\phi \phi } &=&\sin ^{2}\theta \left[\Sigma -a^{2}\left( f(r,\theta)-2\right) \sin ^{2}\theta\right].
\end{eqnarray}%
The metric function $f(r,\theta)$ is given by
\begin{equation}
f(r,\theta) = 1 - \frac{2Mr\sqrt{\Sigma}}{(\Sigma + Q^{2})^{3/2}} + \frac{Q^{2}\Sigma}{(\Sigma + Q^{2})^{2}},
\end{equation}%
with%
\begin{equation}
\Delta_{\text{AGBH}} = \Sigma f(r,\theta)+a^{2}\sin^{2}{\theta},\quad \Sigma = r^{2}+a^{2}\cos ^{2}\theta.
\end{equation}

In order to calculate the difference in the TOA, we need to calculate the null trajectory for the spinning AGBH (8-15) on the equatorial plane, $\theta = \pi/2$, whence $\sin{\theta} = 1$, $\cos{\theta} = 0$ and the function $f(r, \theta)$ reduces to a simple form [actually $A(r)$, just incidentally]:
\begin{equation}
f(r)=1-\frac{2Mr^{2}}{(r^{2}+Q^{2})^{3/2}}+\frac{Q^{2}r^{2}}{(r^{2}+Q^{2})^{2}}.
\end{equation}

The metric (8-15) does not reduce to KNBH, except under the weak field expansion, where $\frac{M}{r}, \frac{Q}{r}<<1$ in the equatorial plane ($\theta = \pi/2$). However, it reduces exactly to KBH at $Q=0$. So we separately write out the exact KNBH metric for calculating the TOA:
\begin{eqnarray}
g_{tt} &=&\frac{a^{2}\sin ^{2}\theta -\Delta _{\text{KNBH}}}{\Sigma}, \\
g_{t\varphi} &=& \frac{a\sin ^{2}\theta \left[ \Delta _{\text{KNBH}}-(r^{2}+a^{2})\right] }{\Sigma}, \\
g_{rr} &=& \frac{\Sigma }{\Delta _{\text{KNBH}}}, \\
g_{\theta\theta} &=& \Sigma, \\
g_{\phi\phi} &=& \frac{\sin ^{2}\theta }{\Sigma }\left[(r^{2}+a^{2})^{2}-%
\Delta_{\text{KNBH}}a^{2}\sin ^{2}\theta\right], \\
\Delta_{\text{KNBH}} &=& r^{2}-2Mr+Q^{2}+a^{2}.
\end{eqnarray}

\section{The difference in the TOA}

To derive the equation for the difference in the TOA, consider a null trajectory $d\tau^{2} = 0$ on the equatorial plane ($\theta = \pi/2$) in the generic metric (8) so that the coordinate time required for light rays along an infinitesimal null world line is given by
\begin{equation}
dt_{\pm} = \frac{d\phi}{g_{tt}}[-g_{t\phi}\pm h(r,\phi)],
\end{equation}%
where
\begin{equation}
h(r,\phi )\equiv \sqrt{g_{t\phi }^{2}-g_{tt}\left\{ g_{rr}\left( \frac{dr}{d\phi }\right) ^{2}+g_{\phi \phi }^{2}\right\}}.
\end{equation}%
We assume the passage of coordinate time to be positive for both $\pm$ sides of the lens. Hence we identify $d\phi >0$ for light rays passing the lens by the co-rotating side ($+$) and $d\phi <0$ for the counter-rotating side ($-$), so that $dt_{+}$ and $dt_{-}$ are both positive. The net difference between the two null rays in the time of arrival (TOA) at the observer is also positive and is given by:
\begin{eqnarray}
dt=dt_{-}-dt_{+}&=&\frac{\left\vert d\phi \right\vert }{g_{tt}}[g_{t\phi}+h(r,\phi )]-\frac{\left\vert d\phi \right\vert }{g_{tt}}[-g_{t\phi}+h(r,\phi )] \nonumber \\
&=& \left\vert \frac{2g_{t\phi }}{g_{tt}}\right\vert \left\vert d\phi \right\vert.
\end{eqnarray}%
The delay $dt$ is due to the frame-dragging effect characterized by $\left(\frac{2g_{t\phi}}{g_{tt}}\right)$, which we are going to compute in this paper. When the lens is not spinning, $g_{t\phi} = 0$, the path lengths of the light ray on both sides of the lens would be the equal and there would be no difference in TOA at the observer, $dt = 0$. Only when the lens is spinning the path lengths will differ -- longer for counter-rotating and shorter for co-rotating sides -- giving rise to the TOA, $dt>0$. We assume that the source, spinning lens and the observer are aligned, that is, they are situated on a straight line (see Fig.1).

\begin{figure}[!t]
  \centerline{\includegraphics[scale=0.7]{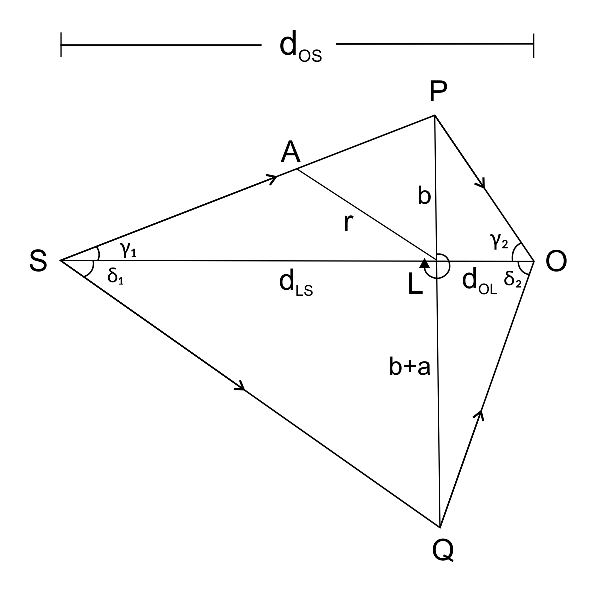}}
  \caption{The generic thin-lens slender quadrilateral (here angles are exaggerated). $S$, $L$ and $O$ are the source, lens and observer respectively aligned on a straight line, $b$ is the impact parameter and $a$ is the spin with axis perpendicular to the paper. The arbitrary angles (32-34) in Sec.4 are as shown.}
\end{figure}

The rays on the equatorial plane are required to pass through the weak field so that the thin-lens approximation is valid, that is, the distance of closest approach $r >> r_{ph}^{\pm}$ on either side of the lens. This large $r$ ensures that, for a given lens of mass $M$ and spin $a$, the quantity $\left(\frac{M}{r}\right) << 1$. We shall apply the generic formula (25) to AGBH by expanding $\left(\frac{2g_{t\phi}}{g_{tt}}\right)$ such that, to first order in $a$ and up to third PPN order in $\left(M/r\right)$, we obtain:
\begin{equation}
dt^{\text{AGBH}} = |d\phi|\left(\frac{1}{c}\right)\left[\frac{4aM}{r}\left\{1 + \frac{2M}{r}\lambda_{1} + \frac{M^{2}}{2r^{2}}\lambda_{2} + ...\right\} \right],
\end{equation}%
where%
\begin{equation}
\lambda_{1} = 1-\frac{Q^{2}}{4M^{2}}, \quad \lambda_{2} = 8-\frac{7Q^{2}}{M^{2}}.
\end{equation}%
For the KNBH, likewise we obtain
\begin{equation}
dt^{\text{KNBH}} = |d\phi|\left(\frac{1}{c}\right)\left[\frac{4aM}{r}\left\{1 + \frac{2M}{r}\xi_{1} + \frac{M^{2}}{2r^{2}}\xi_{2} + ...\right\}\right],
\end{equation}%
where the deviation parameters are%
\begin{equation}
\xi_{1}=1-\frac{Q^{2}}{4M^{2}}, \quad \xi_{2}=8-\frac{4Q^{2}}{M^{2}}.
\end{equation}%
KBH values ($Q=0$) are retrieved when the deviation parameters assume values $\lambda_{1} = \xi_{1} = 1$, $\lambda_{2} = \xi_{2} = 8$. Note that the factor $\left(\frac{4aM}{r}\right)$ multiplying the curly brackets in Eqs.(26,28) need not in general be small due to the presence of spin $a\sim M$. However, it will be seen that practical values associated with the binaries ensure that the terms $a\left(\frac{M}{r}\right)^{2}$ and $a\left(\frac{M}{r}\right)^{3}$ decay very rapidly making the series within the square brackets in (26,28) converge, leaving the leading order term $\left(\frac{4aM}{r}\right) $ to be the main contributor to $dt$.

The total dfference in the TOA $\Delta t^{\text{AGBH}}$ between two null rays traveling from the source to observer along two opposite sides of the intervening spinning lens is%
\begin{eqnarray}
\Delta t^{\text{AGBH}} &=& \left(\frac{1}{c}\right)\int\limits_{0}^{\pi}d\phi\left[\frac{4aM}{r}\left\{1 + \frac{2M}{r}\lambda_{1} + \frac{M^{2}}{2r^{2}}\lambda_{2} + ...\right\}\right] \\
&\equiv& \frac{1}{c}\left(I_{1} + I_{2} + I_{3}\right) = \Delta t_{1} + \Delta t_{2} + \Delta t_{3}.
\end{eqnarray}%
We shall compute the integral locating the spinning lens at the origin of a polar system of coordinates on the equatorial plane ($\theta = \pi/2$). As can be seen, $\left(\frac{Q^{2}}{M^{2}}\right)$ does not influence the leading first order term $\Delta t_{1}^{\text{AGBH}}$. In the following, we shall derive explicit expressions for $\Delta t_{1}^{\text{AGBH}}$, $\Delta t_{2}^{\text{AGBH}}$ and $\Delta t_{3}^{\text{AGBH}}$ within the thin-lens approximation.

\section{Thin-lens approximation}

In realistic lensing configurations, the radius over which bending takes place is of the order of Schwarzschild radius, which is much smaller than the typical distances $d_{OL}$, $d_{OS}$, $d_{LS}$ over which light propagates. Then, to an excellent approximation, light rays can be assumed to propagate as straight lines when far away from the lens, with the bending taking place only at the point lens (Hartle 2003). In a spinning lens situation, we argue that the approximation requires three additional non-trivial conditions to be satisfied as enumerated below.

(i) The first condition is that the rays emerging from $S$, after passing along line segments on either side of spinning lens $L$ in the equatorial plane, should meet exactly at $O$ making a quadrilateral $SPOQ$ of Fig.1. The angles are related by%
\begin{eqnarray}
\gamma_{1} + \gamma_{2} + \measuredangle OPS &=& \pi = \measuredangle OPS + \theta_{1} \nonumber \\
&\Rightarrow& \gamma_{1} + \gamma_{2} = \theta_{1}, \\
\delta_{1} + \delta_{2} + \measuredangle OPS &=& \pi = \measuredangle OPS + \theta_{2} \nonumber \\
&\Rightarrow& \delta _{1}+\delta _{2}=\theta _{2}.
\end{eqnarray}%
In the thin-lens approximation, the relevant angles given below
\begin{equation}
\gamma_{1} = \frac{b}{d_{LS}},\; \gamma_{2} = \frac{b}{d_{OL}},\; \delta_{1} = \frac{b+x}{d_{LS}},\; \delta_{2}=\frac{b+x}{d_{OL}},
\end{equation}%
should be small, where $b$ is the impact parameter, $x(a,b)$ is an unknown function to be determined. For a given $d_{OL}$, the ratio $\chi$ ($=\frac{d_{LS}}{d_{OL}}$) should be such that the angles $\gamma_{1}$, $\gamma_{2}$, $\delta_{1}$, $\delta_{2} << 1$.

Following the Boyer-Lindquist formula (Boyer \& Lindquist 1967) for the bending of light at $P$ and $Q$ respectively, we have%
\begin{eqnarray}
\theta _{1} &=&\frac{4M}{b}\left( 1-\frac{a}{b}\right), \quad \text{(prograde
motion of light)}, \\
\theta _{2} &=&\frac{4M}{b}\left( 1+\frac{a}{b+x}\right), \quad \text{(retrograde
motion of light)}.
\end{eqnarray}%
Neglecting $\left( \frac{a}{b}\right) ^{2}$ and higher orders, we get from the quadrilateral geometry, the root $x=a$, yielding two impact parameters $b$ and $b+a$ on either side as shown in Fig.1.

(ii) The second condition is that the light rays should pass far away from the spinning lens that their trajectories can be approximated by straight lines. To determine how far is far, we need to know the radii of the photon spheres $r_{\text{ph}}^{\pm}$ appearing respectively on the co-rotating ($+$) and counter-rotating sides ($-$) of the lens. For our purposes, we can safely take the Kerr value $r_{\text{ph}}^{\pm} = 2M\left[1 + \cos\left\{\frac{2}{3}\arccos\left(\frac{\mp a}{M}\right)\right\}\right]$ giving an approximate idea of its location and assume the light ray to be traveling in the weak field region at an impact parameter far larger than the photon sphere.

(iii) Thin lens approximation breaks down when $\frac{M}{b} \sim O\left(1\right)$. Therefore, to avoid it, the third condition is that the smaller of the two impact parameters must far exceed the larger of the two radii $r_{\text{ph}}^{\pm}$ of photon spheres, viz., $b >> r_{\text{ph}}^{-}$ or let $b = 10^{n}r_{\text{ph}}^{-}$, where $n > 1$ is any real number. Our idea is to march $b$ towards $r_{\text{ph}}^{-}$ to the extent that the rays preserve the smallness of the angles in Fig.1 as discussed in (i). This algorithm will be exercised to produce Tables 1 and 2 below.

Returning to Fig.1, we have by construction $d_{LS} = \chi d_{OL}$, where $\chi > 0$ is a finite constant, $PLQ \perp OLS$, and the arbitrary angles are as indicated. The radial distance is measured from the lens $L$. By piecewise integration of the polar straight lines in the counter-rotating $\left(\frac{1}{r_{\text{cou}}}\right)$ and co-rotating $\left(\frac{1}{r_{\text{cor}}}\right)$ sectors, we straightforwardly derive the final result by subtracting the integrals over the path lengths, viz., $SQO-SPO$:
\begin{eqnarray}
I_{1} &=&4aM\left[ \int\limits_{-\pi /2}^{-\pi }\frac{1}{r_{\text{cou}}}%
d\phi +\int\limits_{0}^{-\pi /2}\frac{1}{r_{\text{cou}}}d\phi \right] \\
&&-4aM\left[ \int\limits_{\pi /2}^{\pi }\frac{1}{r_{\text{cor}}}d\phi
+\int\limits_{0}^{\pi /2}\frac{1}{r_{\text{cor}}}d\phi \right]
\end{eqnarray}%
\begin{equation}
\Rightarrow \Delta t_{1}^{\text{AGBH}}=\frac{I_{1}}{c}=\frac{8aM\{\chi
\lbrack a(b+d_{OL})+b(b+2d_{OL})]-b(a+b)\}}{b(a+b)c\chi d_{OL}}.
\end{equation}%
By taking the limit where the source and observer are both infinite distance away from the lens, that is when $d_{OL}\rightarrow\infty$, which also implies $d_{LS}\rightarrow\infty$, we have
\begin{equation}
\Delta t_{1}^{\text{AGBH}} = \frac{8aM(a+2b)}{cb(a+b)}.
\end{equation}%
The requirement of thin-lens approximation further implies that $(a/b) << 1$ such that orders of $(a/b)^{2}$ and higher can be neglected, then we end up with
\begin{equation}
\Delta t_{1}^{\text{AGBH}} \simeq \frac{16aM}{cb},
\end{equation}%
which is independent of $Q$ and is precisely the leading order delay obtained by Laguna \& Wolszczan (1997) (See also footnote 2). Thus Eq.(39) generalizes the zeroth order TOA (41) to realistic \textit{finite distance thin-lens geometry}.

In the same way, we can calculate the integrals $I_{2}$ and $I_{3}$ which, to leading orders in $\left(\frac{M}{b}\right)^{2}$ and $\left(\frac{M}{b}\right)^{3}$, work out to
\begin{eqnarray}
\Delta t_{2}^{\text{AGBH}} = \frac{I_{2}}{c} &=& \frac{4aM^{2}}{A_{1}c\left(\chi d_{OL}\right)^{2}} \{A_{1}\pi + A_{2}\chi d_{OL} (\chi - 1) \nonumber \\
&& + (A_{1}+A_{3}d_{OL}^{2})\pi\chi^{2}\}\lambda _{1}, \\
\Delta t_{3}^{\text{AGBH}} = \frac{I_{3}}{c} &=& \frac{2aM^{3}\left[4 - 4/\chi^{3} + B_{1} + B_{2} + B_{3}\right]}{3cd_{OL}^{3}}\lambda _{2},
\end{eqnarray}%
where%
\begin{eqnarray}
A_{1} &=& b^{2}(a+b)^{2}, \\
A_{2} &=& 2b(a+b)(a+2b), \\
A_{3} &=& a^{2}+2ab+2b^{2}, \\
B_{1} &=& \frac{4d_{OL}^{3}}{b^{3}}+\frac{4d_{OL}^{3}}{(a+b)^{3}}, \\
B_{2} &=& \frac{3d_{OL}^{2}(\chi -1)}{\chi b^{2}}+\frac{3d_{OL}^{2}(\chi -1)}{\chi (a+b)^{2}}, \\
B_{3} &=& \frac{3d_{OL}^{2}(\chi ^{2}+1)}{\chi b^{2}}+\frac{3d_{OL}(\chi^{2}+1)}{\chi ^{2}(a+b)},
\end{eqnarray}

The Eqs. (39, 42-49) are the resulting TOA components to be used. For comparison purpose, we will compute the components for the well known KNBH, which turn out to be
\begin{eqnarray}
\Delta t_{1}^{\text{KNBH}} &=&\frac{8aM\{\chi \lbrack
a(b+d_{OL})+b(b+2d_{OL})]-b(a+b)\}}{b(a+b)c\chi d_{OL}} \\
\Delta t_{2}^{\text{KNBH}} &=&\frac{4aM^{2}}{A_{1}c\left(\chi d_{OL}\right)^{2}} \{A_{1}\pi + A_{2}\chi d_{OL} (\chi - 1) \nonumber \\
&& + (A_{1}+A_{3}d_{OL}^{2})\pi\chi^{2}\}\xi _{1}, \\
\Delta t_{3}^{\text{KNBH}} &=&\frac{2aM^{3}\left[ 4-4/\chi
^{3}+B_{1}+B_{2}+B_{3}\right] }{3cd_{OL}^{3}}\xi _{2},
\end{eqnarray}%
where%
\begin{equation}
\xi_{1} = 1-\frac{Q^{2}}{4M^{2}}, \quad \xi_{2} = 8-\frac{4Q^{2}}{M^{2}}.
\end{equation}

Same as in AGBH, the first order effect, $\Delta t_{1}^{\text{AGBH}} = \Delta t_{1}^{\text{KNBH}}$, neither depends on $\left(Q/M\right)$. Since $\lambda_{1} = \xi_{1}$, it is evident also that $\Delta t_{2}^{\text{AGBH}} = \Delta t_{2}^{\text{KNBH}}$, so the difference in TOA between AGBH and KNBH will appear only in the third order term $\Delta t_{3}$ onwards. For two astrophysical binaries, we shall plug the lens parameter values $a$, $M$, distance values $b$, $d_{OL}$, $\chi = d_{LS}/d_{OL}$ into the above equations, take care to preserve the smallness of the angles $b/d_{OL}$, $b/d_{LS} << 1$, and tabulate the components $\Delta t_{1}$, $\Delta t_{2}$ and $\Delta t_{3}$, the last two containing the factor $Q/M$ (Tables 1,2).

\section{Numerical estimates}

\begin{table*}
\caption{The table shows some typical values of components $\Delta t_{1},$ $\Delta
t_{2},$ $\Delta t_{3}$ for the PSR-Cygnus X-1 binary. The last two columns
contain the effect of the gyromagnetic ratio $\left( Q/M\right)$. The
distances $d_{LS}$, $d_{OL}$ in Fig.1 are such that the angles remain small:
$\gamma _{1}\simeq $ tan$\gamma _{1}=b/d_{LS}\simeq \delta _{1}\simeq $ tan$%
\delta _{1}<<1$ etc. The KBH values are given at $Q=0$ (first row). Thus,
even when the impact parameter $b$ is ten thousand times farther than $r_{%
\text{ph}}^{+}$, first column of the table shows that the TOA component $%
\Delta t_{1}$ would be at the $\sim 0.03\mu $sec level that could be
measurable in near future. The other components in the table that contain $%
\left( Q/M\right) $ are $\Delta t_{2},\Delta t_{3}$ but unfortunately they
hold no promise to be measurable even in the far future apart from serving
academic interest.}
\begin{tabular}{|c|c|c|c|c|c|}
\hline
$Q/M$ & $\chi $ & $\Delta t_{1}^{\text{AGBH,KNBH}}$($\mu $sec) & $\Delta
t_{2}^{\text{AGBH,KNBH}}$ ($\mu $sec) & $\Delta t_{3}^{\text{AGBH}}$ ($\mu $%
sec) & $\Delta t_{3}^{\text{KNBH}}$ ($\mu $sec) \\ \hline
$0$ (KBH) & $0.1$ & $0.02825$ & $1.1218\times 10^{-6}$ & $4.8147\times
10^{-11}$ & $4.8147\times 10^{-11}$ \\
$0.5$ & $0.1$ & $0.02825$ & $1.0517\times 10^{-6}$ & $3.7615\times 10^{-11}$
& $4.2129\times 10^{-11}$ \\
$1$ & $0.1$ & $0.02825$ & $0.8413\times 10^{-6}$ & $0.6018\times 10^{-11}$ &
$2.4073\times 10^{-11}$ \\ \hline
\end{tabular}
\end{table*}

Pulsar-BH (PSR-BH) binary systems provide the best laboratory for testing the difference in TOA predictions since variable sources like pulsars, quasars, GRBs etc can give out signals at the instant they are are behind the spinning black hole on the optical axis $OL$ and their times of arrival $\Delta t$ can be measured at the observer $O$. Though a concrete example of such a binary is yet to be detected, the prospects for discovery seem quite promising (Laguna \& Wolszczan 1997). We assume the pulsar orbit to be in the equatorial plane of the spinning lens and the line of sight is perpendicular to the axis of the spin.

\begin{center}
\textit{(a) PSR-Cygnus X-1 binary}
\end{center}

An early estimate was that, of all pulsars discovered so far, a small but significant number of them belong to a PSR-BH category with a BH having masses a few times of solar masses (Hartle 2003). We consider a typical illustration, namely, of a PSR-Cygnus X-1 binary with $M = 14.8M_{\odot} = 2.19\times 10^{6}$ cm, $a = 0.95M = 2.08 \times 10^{6}$ cm (Gou et al. 2011), $d_{OL}=1.86$ kpc $=5.74\times 10^{21}$ cm (Reid et al. 2011). The KBH case corresponds to $Q = 0$, $\lambda_{1} = 1$, $\lambda_{2} = 8$. The numerical value of $r_{\text{ph}}^{\pm}$ for the spinning AGBH is practically insensitive to the small values of $\left(Q/M\right)$ (Toshmatov et al. 2014; Ahmed, Amir \& Ghosh 2019). The Kerr photon sphere radii should suffice, which are $r_{\text{ph}}^{-} = 8.72\times 10^{6}$ cm and $r_{\text{ph}}^{+} = 3.05\times 10^{6}$ cm, so $r_{\text{ph}}^{-}$ is the larger of the two radii. Accordingly, to preserve the thin-lens and PPN approximation, we choose, according to the stipulation Sec.4 (iii), $b = 10^{4} r_{\text{ph}}^{-} = 8.72 \times 10^{10}$ cm, so that $\frac{M}{b} \sim 10^{-5}$ justifying the PPN expansion in the curly bracket in Eqs.(26,27). Therefore, integrands rapidly converge. The lens-source distances $d_{LS} = \chi d_{OL}$ in Fig.1 are varied by varying $\chi$ but preserving the required smallness of the angles (in radian): $\gamma_{1} \simeq b/d_{LS}$, $\delta_{1} \simeq (a+b)/d_{LS}$, $\gamma_{2} \simeq b/d_{OL}$, $\delta_{2} \simeq (a+b)/d_{OL}$.

\begin{center}
\textit{(b) PSR-SgrA* binary}
\end{center}

Recent observations suggest that there are probably $\sim 100$ pulsars surrounding the supermassive spinning BH SgrA* with orbital periods $\lesssim 10$ years (Pfahl \& Loeb 2004) and a few among them are expected to form PSR-BH binaries with stellar sized BH companions residing within $\sim 1$ parsec of SgrA* (Faucher-Giguere \& Loeb 2011). We shall assume the possibility that some of the pulsars cross the optical axis $OLS$ behind SgrA* making a PSR-SgrA* binary.

\begin{table*}
\caption{The table shows some typical components $\Delta t_{1},$ $\Delta t_{2},$ $%
\Delta t_{3}$ for SgrA* with values given by Kato et al. (2010), viz.,
$M=4.2\times 10^{6}M_{\odot }$, $d_{OL}=7.6$ kpc and a \textit{unique} $%
a=0.44M$ so that $r_{\text{ph}}^{-}=2.15\times 10^{12}$ cm, $r_{\text{ph}%
}^{+}=1.51\times 10^{12}$ cm, $b=10^{7}r_{\text{ph}}^{-}$ so that $M/b<<1$.
The distances $d_{LS}$, $d_{OL}$ in Fig.1 are such that the angles remain
small: $\gamma _{1}\simeq $ tan$\gamma _{1}=b/d_{LS}\simeq \delta _{1}\simeq
$ tan$\delta _{1}$ etc. The KBH values are at $Q=0$ (first row).}
\begin{tabular}{|c|c|c|c|c|c|}
\hline
$Q/M$ & $\chi $ & $\Delta t_{1}^{\text{AGBH,KNBH}}$($\mu $sec) & $\Delta
t_{2}^{\text{AGBH,KNBH}}$ ($\mu $sec) & $\Delta t_{3}^{\text{AGBH}}$ ($\mu $%
sec) & $\Delta t_{3}^{\text{KNBH}}$ ($\mu $sec) \\ \hline
$0$ (KBH) & $0.1$ & $4.2117$ & $1.9032\times 10^{-7}$ & $9.2965\times
10^{-15}$ & $9.2965\times 10^{-15}$ \\
$0.5$ & $0.1$ & $4.2117$ & $1.7842\times 10^{-7}$ & $7.2629\times 10^{-15}$
& $8.1345\times 10^{-15}$ \\
$1$ & $0.1$ & $4.2117$ & $1.4274\times 10^{-7}$ & $1.1620\times 10^{-15}$ & $%
4.6483\times 10^{-15}$ \\ \hline
\end{tabular}
\end{table*}

We find that $\Delta t_{1}$ $\sim 4.2$ $\mu$sec allowed by the thin-lens approximation, which should be measurable with current technology provided an appropriate pulsar is identified in the future missions.

\vspace{-2em}
\section{Conclusion}

Among the three categories of BHs, the AGBH is everywhere regular while the other two (KNBH and KBH) are centrally singular, all possessing significantly different qualitative behavior especially in the light cone structure and global properties. Such geometric differences stem from the fact that AGBH is sourced by nonlinear eletrodynamics, while KNBH is sourced by linear Maxwell field and KBH is a vacuum solution. One would naturally like to know if they differ from the observational point of view as well. To address the query, we chose a novel diagnostic, the difference in the TOA, which samples the frame dragging effect of the spinning lens, and calculated the difference assuming the spinning lens partner in the binary to be the three BHs in succession. The generic formula for this purpose is Eq.(25) which resulted in detailed Eqs.(39, 42-53) under thin-lens approximation. The approximation implies that the TOA considered here is a weak field effect. They nevertheless allowed estimation of the effect of the gyromagnetic ratio $\left(Q/M\right)$ on the TOA. The results are tabulated in Tables 1 \& 2. To our knowledge, such a comparative study using generic TOA formula (25) has not been undertaken heretofore.

We wish to once again emphasize that the difference in the TOA is \textit{not} the Shapiro gravitational time delay -- they are fundamentally different physical effects. Enough attention has somehow not been paid to this important distinction, which is clearly explained in the introduction and specifically in footnote 2.

The method adopted in this paper can be applied with ease to any binary system assuming that the source, lens and observer are aligned. An added advantage offered by the formulas (39, 42-53) is that the they take into account the \textit{finite distance thin-lens geometry}, which truly describes realistic astrophysical lens configuration. The resulting order of magnitude estimates for $\Delta t_{1}$, $\Delta t_{2}$ and $\Delta t_{3}$ from Eqs.(39, 42-53) are tabulated for the three illustrative binary systems. They are quite robust, that is, the order of magnitudes remain unaltered even when the observer-lens distance indicator $\chi$ and the gyromagneytic ratio $\left(Q/M\right)$ are varied at will preserving small angles respecting the thin-lens approximation. However, these estimates are to be taken as only suggestive, since no binary system has yet been conclusively identified, although Monte Carlo simulations indicate that the number of PSR--BH binaries could be significant with spinning BH companion having a few solar masses (Gou et al. 2011). We have considered two binary systems, PSR-Cygnus-X1 and PSR-SgrA* and the zeroth order delays have been found to be at the $\mu$sec level quite consistent with similar predictions in the literature (Laguna \& Wolszczan 1997).

The prediction of $\Delta t_{1}$ based on Eq.(39) for PSR-Cygnus-X1 system is about $0.028$ $\mu $sec (Table 1). Achieving the required level of accuracy could be possible in the near future since a precision of $0.1$ $\mu$sec was achieved by (Van Straten et al. 2001) for $PSRJ0437-4715$, a bright millisecond pulsar in a White Dwarf-Neutron Star (WD-NS) binary system. However, the measurement of higher order terms $\Delta t_{2}$ and $\Delta t_{3}$ that contain $\left(Q/M\right)$ would require better than pico-second level accuracy, which is unlikely to be attained even in the far future. As to the PSR-SgrA* binary, it is found that $\Delta t_{1} \sim 4.11$ $\mu$sec (Table 2), which should be measurable provided a suitable variable source is detected from among the pulsars orbiting SgrA* (Faucher-Giguere \& Loeb 2011; Kato et al. 2010) and other complications are taken care of.

Since the effect of $\left(Q/M\right)$ on the difference in TOA is rather tiny in the ideal situation considered above, it may be that TOA is not the best diagnostic. Alternatively, we might think that the predictions might be improved by considering the arrival times at $O$ along the exact null trajectories \textit{grazing} the BH photon spheres $r_{\text{ph}}^{\pm}$ like one does in the case of computing lensing observables in the strong field limit (see, e.g., (Nandi et al. 2017, 2018; Izmailov et al. 2019b, 2019c)). In this case, of course, the weak field thin-lens approximation will break down although it provides a "simple and elegant description of many realistic lensing situations" (Hartle 2003). The strong field limit has its share of woes, one of which is the following: To calculate exact null path lengths, one needs to know the exact bending, but unfortunately the latter reveals a virulent \textit{logarithmic divergence} at the photon sphere preventing a Taylor expansion around it. One way to tackle this problem could be to consider a different expansion, the so called "affine perturbation series", that has been claimed to work up to $1\%$ accuracy to the exact value (Iyer \& Petters 2007; Iyer \& Hansen 2009). We shall deal with these issues in a future work.

For completeness, we comment on some subtleties associated with light propagation in nonlinear electromagnetic field. We emphasize that in our work we adhered strictly to Einstein's theory of gravity. Thus the non-linear electrodynamics form the stress tensor on the right hand side of Einstein's field equations resulting in the exact solution of AGBH, which \textit{subsumes} the electromagnetic nonlinearity in its curved geometry. Light follows null geodesics of that curved spacetime, $d\tau^{2} = 0$, which is what we had considered in accordance with Einstein's theory. The flat background cannot be separated out from the curved geometry due to the universality of gravitation. On the other hand, the work by Novello et al. (2000) suggests a spacetime completely different from that of AGBH. The light propagation in a non-linear electrodynamic field placed on top of a flat \textit{Minkowski background} with metric $\eta^{\mu\nu}$ do not follow geodesics of $\eta^{\mu\nu}$ but can be thought of as following the geodesics of an "effective" curved spacetime $g_{\text{eff}}^{\mu\nu}$ that has nothing to do with the AGBH spacetime. It was already recognized in (Novello et al. 2000) that the analogy between photon propagation in effective spacetime and its behavior in a true gravitational field cannot be pushed too far except merely that in both cases there occurs curved geometries resulting from the corresponding nonlinear processes. There are two entirely different spacetimes arising here -- one is the exact AGBH from Einstein's theory and the other is the effective spacetime perceived by photons alone. Studying TOA in the effective geometry would certainly be an interesting future work.

Note further that the effective metric $g_{\text{eff}}^{\mu\nu}$ for Born-Infeld electrodynamics is based on the flat Minkowski background metric $\eta^{\mu\nu}$ in Euclidean coordinates is given by (Novello et al.2000; Plebansky 1968)
\begin{equation*}
g_{\text{eff}}^{\mu\nu} = \left(b^{2} + \frac{1}{2}F\right) \eta^{\mu\nu} + F_{\lambda}^{\mu} F^{\lambda\nu},
\end{equation*}%
where $b^{2}$ is related to the Born-Infeld electric field parameter. In virtue of the principle of general covariance, one could convert $\eta^{\mu\nu} \rightarrow \gamma^{\mu\nu}$ in non-Euclidean coordinates but that would not introduce any new physics. However, if the effective metric $g_{\text{eff}}^{\mu\nu}$ is based not on a flat background $\gamma^{\mu\nu}$ but on a background metric $g^{\mu\nu}$ for which the curvatures are \textit{non-vanishing}, then $g_{\text{eff}}^{\mu\nu}$ would describe photon propagation in nonlinear electrodynamics placed on top of an already curved background $g^{\mu\nu}$. The work by Jana \& Kar (2015) provides an examplary exercise in this generalized framework applied to the Born-Infeld electrodynamics.

When this article was under review, a very recent work by Toshmatov, Ahmedov \& Malafarina (2021) came to our notice. Using the notion of effective metric, they show that, for any black hole being a charged solution of the field equations of general relativity coupled to the nonlinear electrodynamics, one cannot distinguish the two types of charge (magnetic or electric) through the motion of light rays around it. This is certainly a new and important conclusion that could have far reaching consequences. However, in the present paper, the considered AGBH metric is not an effective metric perceived by the motion of light alone but is an exact solution - its geodesics describe motion of both light and particles alike respecting the universality of gravitation.

\section*{Acknowledgments}

We thank an anonymous reviewer for his/her constructive comments.



\end{document}